# Hill Function-based Model of Transcriptional Response: Impact of Nonspecific Binding and RNAP Interactions

Wenjia Shi[1,a,b], Yao Ma[1,a], Peilin Hu[1], Mi Pang[1], Xiaona Huang[2], Yiting Dang[1], Yuxin Xie[1], Danni Wu[1]

[1] Department of Applied Physics, Xi'an University of Technology, Xi'an, 710054, China.

[2] School of Physics and Electromechanical Engineering, HeXi University, Zhangye, Gansu Province, 734000, China.

[a] These authors contributed equally to this work.

[b] e-mail: wenjia.shi@gmail.com (corresponding author).

## Abstract

Hill function is one of the widely used gene transcription regulation models. Its attribute of fitting may result in a lack of an underlying physical picture, yet the fitting parameters can provide information about biochemical reactions, such as the number of transcription factors (TFs) and the binding energy between regulatory elements. However, it remains unclear when and how much biochemical information can Hill function provide in addition to fitting. Here, started from the interactions between TFs and RNA polymerase during transcription regulation and both of their association-dissociation reactions at specific/nonspecific sites on DNA, the regulatory effect of TFs was deduced as *fold change*. We found that, for weak promoter, *fold change* can degrade into the regulatory factor ($F_{reg}$) which is closely correlated with Hill function. By directly comparing and fitting with Hill function, the fitting parameters and corresponding biochemical reaction parameters in $F_{reg}$ were analyzed and discussed, where the single TF and multiple TFs that with cooperativity and basic logic effects were considered. We concluded the strength of promoter and interactions between TFs determine whether Hill function can reflect the corresponding biochemical information. Our findings highlight the role of Hill function in modeling/fitting for transcriptional regulation, which also benefits the preparation of synthetic regulatory elements.

# 1 Introduction

Gene expression regulation is a very important basic function for biological systems to promote their own adaptation and survivals among external environmental stimuli and the change of their own situations, which occurs at all levels of genetic information transmission, but mainly at the transcription level for prokaryotes. In this paper, we also take prokaryotes as an example to study and discuss. Transcriptional regulation decides whether a gene is expressed or not, and it is an extremely complex portion in gene regulation and is also the current research of interest [1][2][3]. Transcriptional regulations result in the activation or inhibition of specific genes' expression by affecting the binding of RNA polymerase (RNAP) with promoters through the association or dissociation of transcriptional factors (TFs) with promoters, which ultimately participates in various cellular processes, such as embryonic development [4][5]; uptake and utilization of nutrients [6][7]; cell cycles completion [8][9]. TFs are classified as activators and repressors according to their roles in the process of transcription.

Obtaining quantitative gene expression levels based on gene regulation models is so important and challenging for us to understand, predict, and design the function of biological systems. There are many kinds of mathematical models are widely used, such as: Boolean model [10][11], ordinary differential equation (ODE) model [1][15][16], stochastic differential equation model [12][13], Bayesian network model [14]. Among them, ODE model is very popular. As shown in eq. (1), gene transcription regulation can be modeled as an ODE, where $d[X_{\mathrm{down}}]/dt$ is the change rate of the concentration of downstream gene product $X_{\mathrm{down}}$ that is regulated by TF $X$ (we only focus on the regulation of TF, so the degradation of gene product is not considered). This change rate is usually written as the regulatory function $f([X])$, and $[X]$ represents the concentration of the corresponding TF $X$ (activator or repressor). At any time, the concentration of $X_{\mathrm{down}}$ can be determined by $f([X])$. Therefore, $f([X])$ is very critical to accurately describe the regulation of TFs in the ODE models.

$$\frac{d[X_{\text{down}}]}{dt} = f([X]) \quad (1)$$

The regulatory functions based on Hill function is the most widely used due to its simple form and intuition [17][18][19][20][21][22]. At the beginning, Hill function was a phenomenological fitting function proposed by A.V. Hill to describe the binding of oxygen (ligand) with hemoglobin (receptor) at different concentrations [23]. Currently, as a function that can describe the receptor occupancy by ligand, Hill functions are widely used in pharmacology [22], physiology and biochemistry, including the binding of TFs to DNA [1]. Hill function based regulatory function is always represented by eq. (2), where, $\beta$, $K$, $n$, $c$ respectively represents the maximal transcriptional rate, transcriptional threshold, Hill coefficient and basal transcriptional rate. The S-shaped curve of Hill function can describe two response states of gene expression levels as "on" and "off". Such on-off states are an important way for biomolecules or cells to generate efficient responses. Beyond fitting, Hill function in eq. (2) still can be derived through the law of mass action with a simple model considering the association and dissociation of TFs with the specific sites in the promoter as shown in fig. 1A. Particularly, in some certain conditions, we can get the corresponding physical meaning of fitted parameters, where the Hill coefficient $n$ is the number of the TFs binding to promoter, and $K^n$ is the equilibrium dissociation constant of TFs binding to the promoter.

$$f([X]) = \begin{cases} \beta \dfrac{\dfrac{[X]^n}{K^n}}{1+\dfrac{[X]^n}{K^n}} + c & (\text{Hill function for activators}) \\ \beta \dfrac{1}{1+\dfrac{[X]^n}{K^n}} + c & (\text{Hill function for repressors}) \end{cases} \quad (2)$$

However, generally, Hill function is only a fitting function, and the relevant fitting parameters do not have the actual physical meaning described above. There are three important factors: first, Hill coefficient $n$ as an estimated number of TFs binding to the promoter is only accurate under the condition of strong cooperativity between TFs. In more general case, it is not a reliable indicator which can only estimate the lower limit of the real value, and also can be non-integers. Second, the model in fig. 1A only considers the binding of the TFs with the specific sites on the promoter, which leaves out the more essential interactions between TFs and RNAP as well as their interactions with nonspecific binding sites on the DNA chain (shown in fig. 1B). With these considerations, the fitted parameters $n$ and $K^n$ from Hill function in eq. (2) may have relationship with RNAP and nonspecific sites. So, one question is, what are the differences in fitted parameters brought by the second factor we mentioned above? Third, when multiple TFs co-regulate the expression of a gene, logic gates similar to electrical engineering are used to characterize transcriptional regulation effects [25][26]. For example, the fitting effect of AND logic gate can be achieved by multiplying Hill functions of different TFs in eq. (2), and the fitting effect of the OR logic gate can be achieved by adding them together. For convenience, we call the fitting function derived from eq. (2) the "fitting logic gate function" below. Then comes the question: does the parameter fitted by the "fitting logic gate function" have definite and practical physical significance? The first of the above three factors has been clearly studied [27], but the second and third factors are not well understood and are also the two questions to be answered in this paper.

In order to answer the above two questions, in this paper, the regulatory functions were firstly deduced from the binding reactions of TFs, RNAP and DNA sequences (including specific/nonspecific binding sites) based on the law of mass action. We found that the regulatory function is strongly associated with Hill function, so by discussing and analyzing the regulatory function, the first question above (*i.e.* the second factors above) can be answered. Then, by comparing the regulatory functions of two TFs with the classical logic gates (AND, OR), the second question above (*i.e.* the third factors above) can be answered.

Three basic assumptions were set up in this paper: (a) The assumption that RNAP and TFs bind to corresponding specific or nonspecific binding sites on the DNA chain [28]. (b) The assumption of quasi-equilibrium state for time scale separation. That is, the association/dissociation rate between RNAP/TFs and binding sites on the DNA chain is much higher than the transcription rate, and then it is assumed that RNAP's occupation of promoters is in a quasi-equilibrium state [28]. (c) The assumption that the rate of gene expression is proportional to the probability of RNAP occupying promoters [29]. Based on these three assumptions, the probability of RNAP occupying promoters under the influence of TFs was derived in detail from the binding of RNAP and TFs to DNA chains according to the law of mass action. It is concluded that under certain conditions, when there is TF regulation, the ratio of this probability and that without TFs, which is defined as *fold change*, is proportional to a function of TF concentration. This function (hereinafter referred to as the "regulatory factor" and denoted by "$F_{reg}$") is closely related to but fundamentally different from Hill function in eq. (2) and its "fitting logic gate function".

In this paper, the physical process of transcriptional regulation will be described frequently in the form of reaction equations. For convenience, we first agree on the symbolic meanings:

（1）RNAP is expressed as $P$, and TF (activator or inhibitor) is expressed as $X$. The molecule in the free state of any of the above protein molecules $Y$ ($X$ or $P$) is represented as $Y_{\text{free}}$.

（2）$S^Y$ represents the specific binding site of molecule $Y$ on the DNA chain, and $NS$ represents the nonspecific binding site. If there is a protein molecule $Y$ occupying any of the above sites $Z$ ($S^Y$ or $NS$), the protein-site binding state is represented as $YZ$ (*i.e.* $PS^P$ represents the specific binding site on promoters occupied by RNAP, and $XNS$ represents the nonspecific binding site occupied by TF $X$); if site Z is empty, the site in this state is represented as $Z_{\text{free}}$.

（3）$N_Y$ in the mathematical expression represents the number of $Y$, and $[Y]\equiv N_Y/V$ represents the concentration of $Y$ (based on the quasi-equilibrium assumption, defaulted to the concentration after reaching a detailed balance), where $V$ represents the volume of the cell ($V$ is a constant in this paper).

（4）We use $\varepsilon_{Y_1Y_2}$ to represent the interaction energy between two protein molecules $Y_1$ and $Y_2$ at specific binding sites ($Y_1$ and $Y_2$ can represent TFs or RNAP), and the corresponding Boltzmann factors are denoted as $\omega_{Y_1Y_2} \equiv e^{-\varepsilon_{Y_1Y_2}/k_BT}$.

（5）In the time scale of transcriptional regulation, the total concentration of specific binding sites of RNAP in cells can be considered as conserved, denoted as $\eta$.

## 2 Models and results

In this part, we will deduce *fold change* and regulatory factor $F_{reg}$ by using the law of mass action and the definition of physical processes in transcriptional regulation, especially considering the regulations with a single TF and two interacting TFs. After analysis, we found that $F_{reg}$ was associated with Hill function in eq. (2) and its "fitting logic gate function", and then we discussed their differences and connections, focusing in particular on their parameters.

### 2.1 The probability of promoter occupation by RNAP without transcriptional regulation and the definition of regulatory factor

Without the TFs, free RNAP ($P_{\text{free}}$) can bind to specific sites ($S^P_{\text{free}}$) or nonspecific sites ($NS_{\text{free}}$) on DNA chain, this process can be shown by the following reaction equation:

$$S^P_{\text{free}} + P_{\text{free}} \rightleftharpoons PS^P;\, NS_{\text{free}} + P_{\text{free}} \rightleftharpoons PNS \tag{3}$$

The above reactions are considered in detailed balance based on the quasi-equilibrium hypothesis. According to the law of mass action, the equilibrium dissociation constants of specific and nonspecific binding between RNAP and DNA chain can be wrriten as:

$$K^S_{pd} = \left[P_{\text{free}}\right]\left[S^P_{\text{free}}\right] \Big/ \left[PS^P\right]; K^{NS}_{pd} = \left[P_{\text{free}}\right]\left[NS_{\text{free}}\right] \Big/ \left[PNS\right]$$

Meanwhile, the binding energy of RNAP to specific and nonspecific sites on the DNA chain, $\varepsilon^S_{pd}$, $\varepsilon^{NS}_{pd}$ can be expressed as:

$$\varepsilon^S_{pd} = \mu^0_{PS^P} - \mu^0_{S^P_{\text{free}}} - \mu^0_{P_{\text{free}}} = k_BT\ln\left(K^S_{pd} \big/ c_0\right)$$

$$\varepsilon^{NS}_{pd} = \mu^0_{PNS} - \mu^0_{NS_{\text{free}}} - \mu^0_{P_{\text{free}}} = k_BT\ln\left(K^{NS}_{pd} \big/ c_0\right)$$

In all text, $\mu^0_Y$ represents the chemical potential of molecule $Y$, $c_0$ represents the concentration of the referenced state. In this paper, we take $c_0$ as 1 nM. Free RNAP can be regarded as an intermedia te product. As the concentration of free RNAP is very low (RNAP almost binds to specific or nonspecific binding sites on the DNA chain [28], corresponding to assumption (a)), the two reactions of eq. (3) can be written as an equivalent reaction:

$$PNS + S_{\text{free}}^{P} \rightleftharpoons PS^{P} + NS_{\text{free}} \tag{4}$$

It can be seen as a simplification of

$$PNS + S_{\text{free}}^{P} \rightleftharpoons P_{\text{free}} + NS_{\text{free}} + S_{\text{free}}^{P} \rightleftharpoons PS^{P} + NS_{\text{free}}$$

We will use this kind of simplification in this paper, that both RNAP and TFs bind to specific or nonspecific binding sites on the DNA chain, as mentioned in the assumption (a). Assuming that the effective equilibrium dissociation constant of equivalent reaction eq. (4) is $K$, so:

$$k_B T \ln K = \mu_{PS^P}^{0} + \mu_{NS_{\text{free}}}^{0} - \mu_{PNS}^{0} - \mu_{S_{\text{free}}^{P}}^{0} = \varepsilon_{pd}^{S} - \varepsilon_{pd}^{NS}$$

where we defined $\Delta\varepsilon_{pd} \equiv \varepsilon_{pd}^{S} - \varepsilon_{pd}^{NS}$, and we have:

$$K = \exp\left(\Delta\varepsilon_{pd} / k_B T\right) \tag{5}$$

On the other hand, according to the definition of equilibrium dissociation constant, we have:

$$K = \left([PNS]\left[S_{\text{free}}^{P}\right]\right) \Big/ \left(\left[PS^{P}\right][NS_{\text{free}}]\right) = \left(N_{PNS}\left[S_{\text{free}}^{P}\right]\right) \Big/ \left(N_{NS_{\text{free}}}\left[PS^{P}\right]\right)$$

Assuming that $N_{NS} >> N_P$ (*E. coli*'s data supports this hypothesis [30]), thus there is $N_{NS_{\text{free}}} \approx N_{NS}$ and $[NS_{\text{free}}] \approx [NS]$ (it can be regarded as a constant), so $K \approx \left(N_{PNS}\left[S_{\text{free}}^{P}\right]\right) \Big/ \left(N_{NS}\left[PS^{P}\right]\right)$, and then:

$$\left[S_{\text{free}}^{P}\right] \Big/ \left[PS^{P}\right] = \left(N_{NS} e^{\Delta\varepsilon_{pd}/k_B T}\right) \Big/ N_{PNS} \tag{6}$$

Finally, we can get the probability of RNAP occupying the promoter :

$$p_{bound} = \frac{\left[PS^{P}\right]}{\left[PS^{P}\right] + \left[S_{\text{free}}^{P}\right]} = \frac{1}{1 + \dfrac{N_{NS}}{N_{PNS}} e^{\Delta\varepsilon_{pd}/k_B T}} \tag{7}$$

The probability of RNAP occupying the promoter is corrected as $\tilde{p}_{bound}$, shown in eq. (8), when the transcriptional regulation of TFs is considered.

$$\tilde{p}_{bound} = \frac{1}{1 + \frac{N_{NS}}{F_{reg} N_{PNS}} e^{\Delta\varepsilon_{pd}/k_B T}} \tag{8}$$

Among them, $F_{reg}$ is a regulatory factor, that is, due to the presence of TFs, it can be considered that the "effective" number of RNAPs - $N_{PNS}$ that can specifically bind to the promoter is adjusted to $F_{reg}N_{PNS}$. $F_{reg}>1$ with activators; $F_{reg}<1$ with repressors; $F_{reg}=1$ with no TF regulations. In the following, after calculation and analysis, it can be found that in the presence of TFs, $F_{reg}$ will degenerate into Hill function in eq. (2) under certain conditions. Prior to this, the important role of $F_{reg}$ in characterizing gene expression dynamics should be explained.

As assumption (c) stated, the rate of gene expression is proportional to the probability of RNAP occupying promoters. Thus, the effect of TFs on gene expression rate can be characterized by the probability ratio of RNAP occupying promoters with/without TFs. According to eqs. (7-8), the *fold change* of RNAP occupied promoters was defined (*i.e.*, the gene expression rate generated by TFs ).

$$fold\ change \equiv \frac{\tilde{p}_{bound}}{p_{bound}} = \frac{p+1}{p + \frac{1}{F_{reg}}} \tag{9}$$

It can be concluded from the above equation that when $p \equiv \frac{N_{PNS}}{N_{NS}} e^{-\Delta\varepsilon_{pd}/k_B T} \ll \min\left\{1, \frac{1}{F_{reg}}\right\}$, *fold change*$\approx F_{reg}$, that is, under the regulation of TFs, the gene expression rate is proportional to the regulatory factor $F_{reg}$. In this paper, we call it the weak promoter condition; conversely, if this condition is not met, the rate of gene expression under the influence of TFs is also related to the number of RNAP, $\Delta\varepsilon_{pd}$ and the number of nonspecific sites on the DNA chain (namely, the $p$ value in eq. (9)) [31]. Hill function used in this case only has fitting significance, and the related parameters obtained by fitting, such as equilibrium dissociation constant $K^n$ and Hill coefficient $n$ in eq. (2) may have deviated from their original physical meaning. We will conduct some discussions on the influence of $p$ on Hill function fitting in sect. (3)

below.

Next, we will first calculate $F_{reg}$ when there is a single TF, and discuss its difference and connection with Hill function in eq. (2); we will then proceed to calculate the $F_{reg}$ in the presence of two interacting TFs (the case for more TFs is easy to generalize and will not be described here). It shows that the $F_{reg}$ obtained under certain parameter conditions can be fitted with the most general form of Hill function ($n$>1) as shown in eq. (2), and the properties of parameters obtained by fitting will be discussed. Meanwhile, when the weak promoter condition is strictly satisfied ($p$≈0), we compare the $F_{reg}$ with the common "fitting logic gate function", to clarify under what conditions the fitting parameters of this kind of function have underlying significance, and under what conditions they only have fitting significance.

To verify the results we obtained in this section, in sect. (3), a biological system, the regulation of LacI repressor will be fitted by Hill function.

### 2.2 $F_{reg}$ and Hill function in the presence of a single TF

Suppose that a single TF - $X$ is involved in transcriptional regulation process, where $X$ is an activator or inhibitor protein and satisfies $N_X << N_{NS}$ [32]. Based on assumption (a), the concentration of TFs in the free state is very low, that is, TFs are all attached to specific ($S^X$) or nonspecific binding sites ($NS$) on the DNA chain, forming $XS^X$ and $XNS$, which can be transformed into each other and achieve detailed balance in the reaction processes. It should be noted that regardless of whether the TF is at its specific binding site, RNAP has two possibilities: occupying or not occupying the promoter. The only difference is that $XS^X$ specifically bound to DNA attracts or repels RNAP and affects RNAP's occupation of promoters. Let the interaction energy between $X$ and RNAP be $\varepsilon_{XP}$ ($\varepsilon_{XP}$>0 indicates the repulsive interaction, corresponding to a repressor; $\varepsilon_{XP}$<0 indicates attractive interaction, corresponding to an activator; $\varepsilon_{XP}$=0 means no interaction). Therefore, the system has four conformations (Table 1).

Table 1. Four Occupied Conformations with a TF Involved in the Regulation

| Conformational Symbol | $S^X_{\text{free}}S^P_{\text{free}}$ | $XS^X S^P_{\text{free}}$ | $S^X_{\text{free}}PS^P$ | $XS^X PS^P$ |
|---|---|---|---|---|
| Whether $X$ occupies $S^X$ | No | Yes | No | Yes |
| Whether RNAP occupies $S^P$ | No | No | Yes | Yes |

These four conformations can be converted to each other, as shown in fig. 2A, and the corresponding conversion processes can be expressed by the following reaction equations:

Reaction ①: $XNS + S_{\text{free}}^X S_{\text{free}}^P \rightleftharpoons XS^X S_{\text{free}}^P + NS_{\text{free}}$

Reaction ②: $PNS + XS^X S_{\text{free}}^P \rightleftharpoons XS^X PS^P + NS_{\text{free}}$

Reaction ③: $PNS + S_{\text{free}}^X S_{\text{free}}^P \rightleftharpoons S_{\text{free}}^X PS^P + NS_{\text{free}}$

Reaction ④: $XNS + S_{\text{free}}^X PS^P \rightleftharpoons XS^X PS^P + NS_{\text{free}}$

According to the quasi-equilibrium state assumption, the above four reaction processes can be approximated to be in detailed balance. With the law of mass action, the proportional relation between the equilibrium concentrations of the four conformations can be calculated, and then the probability of the promoter being occupied by RNAP.

After calculation (see Appendix A for details), the probability of RNAP occupying the promoter, $\tilde{p}_{bound} = \left(\left[S_{\text{free}} PS^P\right] + \left[XSPS^P\right]\right) / \eta$ has the form shown in eq. (8), where $\eta$ represents the promoter concentration that RNAP can specifically bind, namely, $\eta = \left[S_{\text{free}}^X S_{\text{free}}^P\right] + \left[XSS_{\text{free}}^P\right] + \left[S_{\text{free}} PS^P\right] + \left[XSPS^P\right]$, and $F_{reg}$ is:

$$F_{reg} = \frac{1 + \dfrac{[XNS]}{K_X}\omega_{XP}}{1 + \dfrac{[XNS]}{K_X}} \qquad (10)$$

where, $K_X \equiv [XNS]\left[S_{\text{free}}^X S_{\text{free}}^P\right] / \left[XS^X S_{\text{free}}^P\right]$ represents the effective equilibrium dissociation constant of the binding of the TF to its specific binding site. The effective equilibrium dissociation constant $K_Y$ of molecule $Y$ mentioned later in this paper is completely similar and won't be covered again below.

The number of TFs in the biological cells we are concerned with is much larger than the number of specific binding sites [32][33] [34][35][36], so $[X] \approx [XNS]$. Thus, eq. (10) can be further approximated as:

$$F_{reg}([X]) = \frac{1 + \frac{[X]}{K_X}\omega_{XP}}{1 + \frac{[X]}{K_X}} \tag{11}$$

For the repressor, $0 \le \omega_{XP} < 1$, and eq. (11) can be simplified as:

$$F_{reg}([X]) = (1 - \omega_{XP})\frac{1}{1 + \frac{[X]}{K_X}} + \omega_{XP}$$

The above equation is exactly corresponding to Hill function of repressors in eq. (2) ($1 - \omega_{XP}$ can be absorbed into the maximum transcription rate $\beta$; Hill coefficient $n$ here is equal to 1. More general case with Hill coefficient $n>1$ will be discussed in sect. (2.3.2)), where the second term corresponds to the basal expression term, which is non-ignorable when the repressor concentration is relatively high. If it is a strong repressor, $\omega_{XP}$ is considered to be approximately 0, then $F_{reg}$ can be as follows:

$$F_{reg}([X]) = \frac{1}{1 + \frac{[X]}{K_X}}$$

For the activator $\omega_{XP} > 1$, in eq. (11) can be simplified as:

$$F_{reg}([X]) = (\omega_{XP} - 1)\frac{\frac{[X]}{K_X}}{1 + \frac{[X]}{K_X}} + 1$$

The above equation is exactly corresponding to Hill function of activators in eq. (2) ($\omega_{XP} - 1$ can be absorbed into the maximum transcription rate $\beta$), where the second term, *i.e.*, the basal expression term is non-ignorable when the activator concentration $[X]$ is relatively low. If it is a strong activator with $\omega_{XP} \gg 1$, then when the concentration of activator $[X]$ is finite, $F_{reg}$ can be as follows:

$$F_{reg}([X]) \approx (\omega_{XP}-1)\frac{\frac{[X]}{K_X}}{1+\frac{[X]}{K_X}} \propto \frac{\frac{[X]}{K_X}}{1+\frac{[X]}{K_X}}$$

Through the above calculation and discussion, we obtained probability $\tilde{p}_{bound}$ as eq. (8) and $F_{reg}$ as eq. (11) . We found in the condition of weak promoter, the rate of gene expression affected by TFs is directly proportional to $F_{reg}$, and at that time $F_{reg}$ can degenerate into Hill function ($n$=1) of eq. (2), which at this point is not only a fitting function, but also whose fitting parameter $K$ can reflect the effective equilibrium dissociation constant indicating the binding of the TF to its specific binding site. This result provides a physical basis for Hill function in eq. (2) to be widely used to describe gene expression rates (more general cases with Hill coefficient $n$>1 will be discussed in sect. (2.3.2)).

### 2.3 $F_{reg}$ and Hill functions in the presence of two interacting TFs

#### 2.3.1 Derivation in general case

Suppose that the specific binding site of TF - $X_i$ is $S^{X_i}$ ($i$=1,2), and there are 8 occupied conformations with TFs and RNAP binding at the DNA site (Table 2).

Table 2. Eight Occupied Conformations of the System with Two TFs Involved in the Regulation

| Conformational Symbol | Whether $X_1$ Occupies $S^{X_1}$ | Whether $X_2$ Occupies $S^{X_2}$ | Whether RNAP Occupies $S^P$ |
|---|---|---|---|
| $S_{\text{free}}^{X_1}S_{\text{free}}^{X_2}S_{\text{free}}^{P}$ | No | No | No |
| $X_1S^{X_1}S_{\text{free}}^{X_2}S_{\text{free}}^{P}$ | Yes | No | No |
| $S_{\text{free}}^{X_1}X_2S^{X_2}S_{\text{free}}^{P}$ | No | Yes | No |
| $S_{\text{free}}^{X_1}S_{\text{free}}^{X_2}PS^{P}$ | No | No | Yes |
| $X_1S^{X_1}X_2S^{X_2}S_{\text{free}}^{P}$ | Yes | Yes | No |
| $X_1S^{X_1}S_{\text{free}}^{X_2}PS^{P}$ | Yes | No | Yes |
| $S_{\text{free}}^{X_1} X_2 S^{X_2} PS^P$ | No | Yes | Yes |
| $X_1 S^{X_1} X_2 S^{X_2} PS^P$ | Yes | Yes | Yes |

The eight conformations can be converted to each other by corresponding reactions (fig. 2B). Based on the quasi-equilibrium state assumption, we can solve the proportion of equilibrium concentration of each conformation through the detailed balance conditions of these reactions, and then obtain $\tilde{p}_{bound}$ in this case.

The details of calculation can be seen in Appendix D. Finally, the obtained $\tilde{p}_{bound}$ has the same form as eq. (8), where $F_{reg}$ is:

$$F_{reg} = \frac{1 + \frac{[X_1 NS]\omega_{X_1P}}{K_{X_1}} + \frac{[X_2 NS]\omega_{X_2P}}{K_{X_2}} + \frac{[X_1 NS][X_2 NS]\omega_{X_1P}\omega_{X_2P}\omega_{X_1X_2}}{K_{X_1}K_{X_2}}}{1 + \frac{[X_1 NS]}{K_{X_1}} + \frac{[X_2 NS]}{K_{X_2}} + \frac{[X_1 NS][X_2 NS]\omega_{X_1X_2}}{K_{X_1}K_{X_2}}} \tag{12}$$

$K_{X_i}$ represents the effective equilibrium dissociation constant ($i$=1,2) for binding of $X_i$ to its specific binding site. When the number of TFs is much greater than the number of their specific binding sites, $[X_i] \approx [X_i NS]$. Therefore, eq. (12) can be further approximated as:

$$F_{reg}\left([X_1],[X_2]\right) = \frac{1 + \frac{[X_1]\omega_{X_1P}}{K_{X_1}} + \frac{[X_2]\omega_{X_2P}}{K_{X_2}} + \frac{[X_1][X_2]\omega_{X_1P}\omega_{X_2P}\omega_{X_1X_2}}{K_{X_1}K_{X_2}}}{1 + \frac{[X_1]}{K_{X_1}} + \frac{[X_2]}{K_{X_2}} + \frac{[X_1][X_2]\omega_{X_1X_2}}{K_{X_1}K_{X_2}}} \tag{13}$$

Next, we will explain the case that eq. (13) is a Hill function with cooperative effect under certain parameter conditions, that is, the corresponding eq. (2) with $n$>1.

### 2.3.2 Cooperative regulation and Hill coefficient

The so-called "cooperative regulation" refers to the fact that two TFs $X_1$ and $X_2$ promote (positive cooperativity) or inhibit (negative cooperativity) specific binding of one to DNA because of the specific binding of the other. The most typical positive cooperativity is defined as both TFs bind to their specific binding sites or neither at all. This is because the binding of a single TF to its specific binding site is unstable, but the binding of two TFs at the specific binding site is stabled due to a very strong mutual attraction between them. This situation is equivalent to that in eq. (13), both $K_{X_1}$ and $K_{X_2}$ are extremely large, but $K_{X_1}K_{X_2}/\omega_{X_1X_2}$ is finite size. Therefore, its value is supposed as $\tilde{K}^2$, which is the effective equilibrium dissociation constant of two TFs binding to the promoter sequence. In this way, eq. (13) degenerates into:

$$F_{reg}\left([X_1],[X_2]\right)=\frac{1+\frac{[X_1][X_2]}{\tilde{K}^2}\omega_{X_1P}\omega_{X_2P}}{1+\frac{[X_1][X_2]}{\tilde{K}^2}} \tag{14}$$

If the two TFs are the same protein, that is, $[X_1]=[X_2]\equiv[X],\omega_{X_1P}=\omega_{X_2P}\equiv\omega$, then the above equation can be further written as follows:

$$F_{reg}([X])=\frac{1+\frac{[X]^2}{\tilde{K}^2}\omega^2}{1+\frac{[X]^2}{\tilde{K}^2}}=\begin{cases}\left(\omega^2-1\right)\frac{\frac{[X]^2}{\tilde{K}^2}}{1+\frac{[X]^2}{\tilde{K}^2}}+1 & \left(\text{for activators}\right)\\ \left(1-\omega^2\right)\frac{1}{1+\frac{[X]^2}{\tilde{K}^2}}+\omega^2 & \left(\text{for repressors}\right)\end{cases} \tag{15}$$

Eq. (15) corresponds exactly to Hill function of $n$=2 in eq. (2). More generally, the case that $n$>2 can be obtained in the same way, which will not be repeated here. The larger $n$ reflects the stronger the positive cooperativity effect. From the above derivation, we found that, where there is a strong cooperativity interaction between TFs, Hill coefficient is a more accurate estimate of the number of ligands, otherwise, it is not not a reliable indicator. In the general fitting process, Hill coefficient is usually a non-integer within the range of 1~10 [27]; Meanwhile, the fitted regulatory threshold $K^2$ reflects effective equilibrium dissociation constant of two TFs binding to the promoter.

### 2.4 Implementation of $F_{reg}$ and logical functions

The expression of a gene is usually regulated by multiple TFs, and the interaction between multiple TFs makes the fitting form of Hill function appear in various forms in some parameter range. "Fitting logic gate function" based on Hill function in eq. (2) is a common approach to construct a fitting function. For example, AND logic gate function with the multiplication form of Hill function in eq. (2) and the OR logic gate function with the summation form [37]. Such fitting functions are essentially different from the $F_{reg}$ derived from the underlying reactions, and these differences will be reflected by the fitting parameters. In the following subsections, based on weak promoter, we will focus on the properties of fitting parameters of "fitting logic gate functions" in the case of two co-regulated TFs and two kinds of logic. For convenience, each TF below is regarded as a monomer, which easily results in a fitted Hill coefficient close to 1. We will not disscuss the situations that TFs are polymers with cooperativity among monomers, where the Hill coefficient greater than 1 and the equivalent parameters can be replaced for corresponding TFs in $F_{reg}$ like eq. (13), which is similar to that we did in eq. (14) and also depends on the coopertivity strength.

### 2.4.1. AND logic

When there is no interaction between two TFs - $X_1$ and $X_2$ at specific binding sites, that is, they are regulated independently, $\omega_{X_1X_2} = e^{-\varepsilon_{X_1X_2}/k_BT} = 1$ in eq. (13), $F_{reg}$ is the product of regulatory factors in the presence of a single TF.

$$F_{reg}\left([X_1],[X_2]\right) = \frac{1+\frac{[X_1]}{K_{X_1}}\omega_{X_1P}}{1+\frac{[X_1]}{K_{X_1}}} \frac{1+\frac{[X_2]}{K_{X_2}}\omega_{X_2P}}{1+\frac{[X_2]}{K_{X_2}}} = F_{reg}\left([X_1]\right)F_{reg}\left([X_2]\right) \quad (16)$$

Where, $F_{reg}\left([X_i]\right)$ ($i$ =1,2) represents the regulatory factor when a single TF exists independently, which has the same form as eq. (11), namely:

$$F_{reg}\left([X_i]\right) = \frac{1+\frac{[X_i]}{K_{X_i}}\omega_{X_iP}}{1+\frac{[X_i]}{K_{X_i}}} = \begin{cases} \left(\omega_{X_iP}-1\right)\dfrac{\frac{[X_i]}{K_{X_i}}}{1+\frac{[X_i]}{K_{X_i}}}+1 & \text{(for activators)} \\ \left(1-\omega_{X_iP}\right)\dfrac{1}{1+\frac{[X_i]}{K_{X_i}}}+\omega_{X_iP} & \text{(for repressors)} \end{cases}$$

Eq. (16) combines $F_{reg}([X_i])$ as eq. (11) in the form of multiplication to obtain the joint regulatory factor $F_{reg}([X_1],[X_2])$, which is the AND logic in many application examples [18], typically characterized by obvious regulatory effects only at high concentrations of both two (or more) TFs. The two transcriptional activators MelR and CRP proteins in *E. coli* bind to the melAB promoter by AND logic [38].

Starting from eq. (13) and combining with the parameter constraints discussed above, the effect of AND logic can be obtained. For the regulation of the two activators shown in fig. 3A, the activation effect of AND logic can be obtained by taking $\omega_{X_1P} = \omega_{X_2P} = 20$, $\omega_{X_1X_2} = 1$ and $K_{X_1} = K_{X_2} = 10000\text{nM}$ in eq. (13). The normalized regulatory factor, namely $\tilde{F}_{reg}([X_1],[X_2]) \equiv F_{reg}([X_1],[X_2]) / [F_{reg}([X_1],[X_2])]_{\max}$, is shown in fig. 3A. Simultaneously, the fitting effect of "fitting logic gate function" in eq. (17) formed by product combination of Hill functions as eq. (2) is discussed.

$$f_{\text{AND}}([X_1],[X_2]) = \prod_{i=1}^{2} \left[ \frac{([X_i]/K_i)^{n_i}}{([X_i]/K_i)^{n_i} + 1} + c_i \right] \tag{17}$$

$n_i$, $K_i$ and $c_i$ in eq. (17) respectively represent Hill coefficient, regulatory threshold and basal expression corresponding to $X_i$ ($i$=1,2). Strictly speaking, $c_{\mathrm{i}}$ in eq. (17) corresponds to $c/\beta$ in eq. (2), where the maximal transcriptional rate $\beta_i$ here is extracted as a global factor and disappears after normalization, so we ignore the effects of maximum transcriptional rate. Similar treatment is also used in sect. (2.4.2) and sect. (3). We use the normalized fitting function $\tilde{f}_{\text{AND}}([X_1],[X_2]) \equiv f_{\text{AND}}([X_1],[X_2]) / [f_{\text{AND}}([X_1],[X_2])]_{\max}$ to fit $\tilde{F}_{reg}([X_1],[X_2])$, the results obtained are shown in fig. 3B. The fitting parameters are $n_i$=1, $K_i$=10000nM=$K_{Xi}$, $c_i$=0.0526, and the corresponding root-mean-square (RMES) RMES=9.2570×$10^{-12}$. It is easy to understand why the fitting is so accurate. If $n_i$=1, $K_i$=$K_{Xi}$, $c_i = 1/(\omega_{X_iP} - 1)$ in eq. (17), there is only a constant factor difference between eqs. (16) and (17), which will be strictly equal after normalization, and the basal expression $c_i$=0.0526 fitted above is exactly equal to $1/(\omega_{X_iP} - 1)$ in numerical value. In other words, when the weak promoter condition is met and TFs are regulated independently of each other, combining Hill functions through AND logic (eq. (17)) is a good fitting method for regulatory effects, and the corresponding fitting parameters (such as Hill coefficient $n_i$, regulation threshold $K_i$) can correspond to the actual reaction parameters. This is essentially because $F_{reg}$ can decouple the contribution of each TF through factor decomposition when independent regulation conditions are met among TFs.

### 2.4.2. OR logic

When there is competition between the binding of two TFs $X_1$ and $X_2$ and the specific binding site, especially when the specific binding site of the two TFs overlap with each other, *i.e.*, there is a strong repulsion, it can be obtained that $\omega_{X_1X_2} = e^{-\varepsilon_{X_1X_2}/k_BT} = 0$ in eq. (13), then we have:

$$F_{reg}\left([X_1],[X_2]\right) = \frac{1 + \frac{[X_1]\omega_{X_1P}}{K_{X_1}} + \frac{[X_2]\omega_{X_2P}}{K_{X_2}}}{1 + \frac{[X_1]}{K_{X_1}} + \frac{[X_2]}{K_{X_2}}} \qquad (18)$$

where $X_1$ and $X_2$ can be activators or repressors. Eq. (18) can be used as one of the ways to realize OR logic in gene regulation. As shown in fig. 3C (similar to the processing in sect. (2.4.1), regulatory factor is normalized $\tilde{F}_{reg}([X_1],[X_2]) \equiv F_{reg}([X_1],[X_2]) / [F_{reg}([X_1],[X_2])]_{\max}$ ), the calculation results after taking $\omega_{X_1P} = \omega_{X_2P} = 20$, $\omega_{X_1X_2} = 0$ and $K_{X_1} = K_{X_2} = 200\text{nM}$ show that when either of the two TFs (activators) has a high concentration, an obvious activation effect will occur. This is because the two activators, although competing with each other, readily bind to their respective specific binding sites (as a relatively small number, $K_{X_1} = K_{X_2} = 200\text{nM}$ indicates that the TF readily binds to its specific binding site), and both recruit RNAP to bind to the promoter ( $\omega_{X_1P} = \omega_{X_2P} = 20$ ), so high concentrations of either activator can achieve the activation effect.

It should be noted that in order to implement OR logic, Hill functions in eq. (2) are used as fitting functions in the form of summation in some studies. Taking two activators as examples, the fitting function is shown in eq. (19),

$$f_{\mathrm{OR}}([X_1],[X_2])=\sum_{i=1}^{2}\left[\frac{([X_i]/K_i)^{n_i}}{([X_i]/K_i)^{n_i}+1}+c_i\right] \tag{19}$$

where $n_i$, $K_i$ and $c_i$ represent the Hill coefficient, regulatory threshold and basal expression corresponding to TF $X_i$ ($i$=1,2) respectively. Here we are adding together the two terms equally weighted because in eq. (18) $\omega_{X_1P}=\omega_{X_2P}$ and $K_{X_1}=K_{X_2}$ are set, which ensures that $F_{reg}([X_1],[X_2])$ unchanges after swapping $[X_1]$ and $[X_2]$. This method is simple and intuitive, but its actual fitting effect is not so good (fig. 3D). Similar to the processing method in sect. (2.4.1), the normalized fitting function $\tilde{f}_{\mathrm{OR}}([X_1],[X_2])\equiv f_{\mathrm{OR}}([X_1],[X_2])\big/\left[f_{\mathrm{OR}}([X_1],[X_2])\right]_{\max}$ is used to fit $\tilde{F}_{reg}([X_1],[X_2])$, and the fitting parameters $n_i$=1.6608, $K_i$= 241.75nm, $c_i$ =1.3502, and RMES=0.0459 are obtained. Fig. 3D shows that if eq. (19) is used as a fitting function, not only the fitting accuracy is lower, but also the performance of realizing OR logic is inferior to eq. (18). In fig. 3C, when one of the $X_1$ and $X_2$ levels is low and the other is high, the regulation effect of "on" can be achieved as both levels are high. This is because the maximum value of eq. (18) at this time is $\left[F_{reg}([X_1],[X_2])\right]_{\max}=\max\{\omega_{X_1P},\omega_{X_2P}\}$, and when the level of one of the two TFs, such as $X_1$, is high and the level of $X_2$ is low, it leads $F_{reg}\left([X_1]/K_{X_1}>>1,[X_2]/K_{X_2}<<1\right)$ tends to $\omega_{X_1P}$. In our example, $\omega_{X_1P}=\omega_{X_2P}$, so $F_{reg}\left([X_1]/K_{X_1}>>1,[X_2]/K_{X_2}<<1\right)$ tends to $\left[F_{reg}([X_1],[X_2])\right]_{\max}$. Thus it is easy to achieve a good "on" regulation effect. However, in fig. 3D, when one of the $X_1$ and $X_2$ levels is low and the other is high, the regulation effect of "on" cannot be achieved as shown in fig. 3C, that is, the regulation performance of OR is reduced. This is because for eq. (19), its maximum value $[f_{\mathrm{OR}}([X_1],[X_2])]_{\max}=2+c_1+c_2$, and $f_{\mathrm{OR}}\left([X_1]/K_{X_1}>>1,[X_2]/K_{X_2}<<1\right)$ tends to $1+c_1+c_2$ when the level of one of the two TFs, such as $X_1$ is high and the level of $X_2$ is low. If it is expected to achieve the "on" regulatory effect as good as both levels are high, only $c_1$ (or $c_2$) is much greater than 1, but in this way, it becomes $f_{\mathrm{OR}}([X_1],[X_2])\approx c_1$(or $c_2$), losing the regulatory significance of TFs (at this point, the basal expression term plays a dominant role and is approximately constant).

As can be seen from the above discussion, it is difficult to achieve a good OR logical regulation effect in the fitting process by taking the summation of Hill functions as the fitting function, which is essentially because $F_{reg}$ cannot be decomposed into the form of summation by means similar to that in sect. (2.4.1), so as to achieve the effect of decoupling the contribution of each TF.

## 3 Biological system with Hill function

Finally, we take repressor LacI of lactose operator system, as an example to be fitted by the general form of Hill function in eq. (2), discuss the fitting parameters and their biological significance. In the absence of lactose and related inducers, a dimer formed by LacI monomer and a tetramer formed by two LacI dimers bind to the lactose operon to repress RNAP's occupation of promoters and the expression of lactose metabolism genes. Let R represents LacI dimer as the TF of concern, two R's are considered to bind to promoters to form transcriptional regulation (although only one LacI dimer in the tetramer binds to the promoter sequence in the wild-type condition). *Fold change* and $F_{reg}$ corresponding to this regulatory process can be deduced as eqs. (20-21). Hill function of repressors in eq. (2) is used to fit these two functions, and then the Hill coefficient and regulatory threshold obtained by fitting are observed (since LacI is a strong repressor, the basal expression term is directly set to 0 during the fitting using eq. (2), and the maximal transcriptional rate is a global factor which should be ignored, similar to the statement in sect. (2.4)). Experimental parameters in eqs. (20-21) are obtained from refs. [25] and [39] , and response curves of *fold change*, $F_{reg}$, as well as corresponding Hill function fitting curves as LacI dimer concentration changes are shown in fig. 4A-B.

$$fold\ change = \frac{1+p}{1+p+\frac{2[R]}{K_R}+\frac{[R]^2}{K_R^2}\omega_{RR}} \tag{20}$$

$$F_{reg} = \frac{1}{1+\frac{2[R]}{K_R}+\frac{[R]^2}{K_R^2}\omega_{RR}} \tag{21}$$

As shown in figs. 4A-B, there is a small difference in *fold change* and $F_{reg}$ curve as a function of [$R$] ($p$=0.33 in figs. 4A-B), especially when $\omega_{RR}$ increases (fig. 4B), the difference is not significant in the larger concentration range of [$R$] with the rapid saturation of the response curve. When Hill function is used to fit *fold change* and $F_{reg}$ respectively, it can be seen from fig. 4A or 4B that $p$ only affect $K$ but not $n$. After $\omega_{RR}$ changes, Hill coefficient $n$ and regulatory threshold $K$ are different as shown in figs. 4A and 4B. Meanwhile, Hill coefficient $n$ obtained by fitting quantitatively cannot directly correspond to the number of TFs involved, but when the cooperative factor $\omega_{RR}$ is larger, they are closer. The regulatory threshold $K$ obtained by fitting decreases with the increase of $\omega_{RR}$.

After further changing the values of $p$ and $\omega_{RR}$, Hill function is used to fit the *fold change* curve with [$R$]. A series of Hill coefficients $n$ and regulatory threshold $K$ are obtained. As shown in fig. 4C, the relationship curve between Hill coefficient $n$ and the cooperative factor $\omega_{RR}$ overlaps as $p$ changes, indicating that $n$ has nothing to do with $p$ and its value tends to 2 as $\omega_{RR}$ increases. This feature is consistent with the conclusion in figs. 4A-B, which is intuitively understood that $n$ is only related to the cooperativity between repressors but has nothing to do with RNAP. Fig. 4C shows that Hill coefficient obtained by fitting is an accurate indicator of the number of TFs with strong cooperativity between TFs, which is independent of whether the weak promoter condition is satisfied.

At the same time, we use Hill function of the repressors in eq. (2) to fit *fold change* in eq. (20). Through simple derivation, the regulatory threshold is obtained as: $K = K_R(1+p)^{1/n} g(\omega_{RR})$, where $g(\omega_{RR})$ is only a function of $\omega_{RR}$ (see Appendix E for details). So, $K$ is not only related to $K_R$ (effective equilibrium dissociation constant), but also related to $\omega_{RR}$ and $p$. The solid blue line in fig. 4D shows the variation trend of $g(\omega_{RR})$ with $\omega_{RR}$. When $\omega_{RR}$ is large, the trend tends to decrease in a power law with $g(\omega_{RR}) \approx \omega_{RR}{}^{-1/2}$ (shown by the red dotted line in fig. 4D). When the weak promoter condition is satisfied ($p$≈0), there is $K \approx K_R g(\omega_{RR})$. In other words, with this weak promoter condition, the regulatory threshold obtained by Hill function fitting only reflects the binding energy between TFs and specific sites (*i.e.*, the effective equilibrium dissociation constant $K_R$) and the binding energy between TFs, otherwise, the number of RNAP and RNAP binding energy with DNA will also affect $K$.

## 4 Discussion

Based on the quasi-equilibrium assumption in the transcriptional regulation process of prokaryotic system, with the law of mass action, the regulatory factor $F_{reg}$ of TFs which influences the probability of RNAP occupying of promoter is obtained, namely eqs. (10) and (12), and Hill function related discussion was conducted on this basis. On the other hand, Rob Philips *et al.* proposed a gene transcriptional regulation model based on equilibrium statistical mechanics for prokaryotic systems in ref. [28]. The law of mass action and statistical mechanics of equilibrium are different expressions of equilibrium systems, in which the former is based on thermodynamic theory (macroscopic theory) and the latter is based on statistical ensemble theory (microscopic theory), and they should be self-consistent in nature. In fact, eqs. (10) and (12) obtained by the law of mass action are very close to the results of the statistical mechanical model in ref. [28], and the only difference is that the concentration of TFs in eqs. (10) and (12) corresponds to the concentration at the nonspecific binding site on the DNA chain [$XNS$], which can be interpreted as the concentration of free TFs compared with the TFs at the specific binding site; in ref. [28], the concentration of TFs present in the regulatory factors corresponds to the total concentration of [$X$] (including concentration of TFs specifically and nonspecifically bound to the DNA chain). In other words, the results in ref. [28] are completely consistent with eqs. (11) and (13) in this paper, but with slight difference with (10) and (12). In ref. [28], the authors assume that there is only one specific binding site for both RNAP and TFs in the prokaryotic system, while the law of mass action applies only when the molecules number of each reactant/product is much greater than 1, which deviates from the assumption in ref. [28] and may lead to such subtle differences in the results of the two. However, in this paper, we do not consider such differences to be significant. For example, the typical situation concerned in the assumptions used in the analysis from eqs. (10) to eqs. (11) and from eqs. (12) to eqs. (13) is that the number of TFs is much greater than the number of their specific binding sites [32][33][34][35][36] , so the vast majority of TFs are at nonspecific binding sites, namely $[XNS] \approx [X]$. In this case, the difference is negligible.

## 5 Conclusion

As a widely used mathematical model for transcriptional regulation, Hill function has important characteristics: first, it is simple in form and has S-shaped response characteristics; second, it can reflect the corresponding biochemical reaction image by fitting parameters to a certain extent. But the second point is not always possible, especially when we consider more details of transcriptional regulatory responses, such as RNAP, specific and nonspecific binding sites, which may limit Hill function to only fitting roles. How to mine more information about the underlying reaction from fitting parameters is the focus of this paper, so that we can prepare to synthesis high-precision regulatory elements through the feedback information of fitting parameters, and also provide more theoretical reference for the application of Hill function in gene transcription regulation models.

In this paper, combined with the law of mass action and based on the underlying transcriptional regulatory reactions (including association and dissociation reactions of TFs, RNAP and DNA sequences), the following result is obtained, that is, the effect of TFs on the probability of RNAP occupying promoters in the prokaryotic system, namely *fold change* and regulatory factor $F_{reg}$. By comparing them with Hill function, the fitting parameters of Hill function and their properties are studied. The results show that:

First of all, $F_{reg}$ can reflect the influence of TFs on gene expression changes (*fold change*) only under the condition of weak promoter, and at that time $F_{reg}$ is comparable with Hill function representing the effect of transcriptional regulation. When the weak promoter condition is not met, the influence of TFs on gene expression needs to be characterized by *fold change* instead of $F_{reg}$. At this point, the number of RNAP, the specific and nonspecific binding energy difference between RNAP and DNA sequence (see $p$ in eq. (9)) are all involved in the influence of TFs on gene expression changes.

On the basis of meeting the weak promoter condition, first, taking a single TF involved in regulation as an example, the following results are obtained: $F_{reg}$ in eq. (11) has the same mathematical form as Hill function in eq. (2) (with $n$=1), that is, Hill function in eq. (2) can faithfully reflect the transcription activity under the weak promoter condition. At this point, the regulatory threshold $K$ in eq. (2) will correspond to $K_X$ in eq. (11), which has practical physical significance and corresponds to the effective equilibrium dissociation constant of TF binding to its specific binding site.

In the meantime, we considered the regulations with cooperativity. $F_{reg}$ that two TFs jointly regulate the expression of the same gene, namely eq. (13) is deduced. When the cooperative interaction between TFs is strong, $F_{reg}$ can degenerate into the form of Hill function in eq. (14). Once the two TFs are the same protein, eq. (14) can transform to eq. (15) which corresponds to the situation that $n$>1 in eq. (2). This time, Hill coefficient obtained by fitting is an accurate indication of the number of TFs, and the regulatory threshold obtained by Hill function fitting will provide the information of the effective equilibrium dissociation constant of two TFs binding to the promoter sequence. These results provides eq. (2) a solid modeling power especially in the simutation of weak promoter and strong interactions among monomers.

Besides, from the perspective of gene expression effect, $F_{reg}$ of the two TFs, namely eq. (13), can achieve good AND logical regulation function and relatively poor OR logical regulation function with appropriate parameter values. In the absence of interaction between two TFs, the regulatory threshold $K$ obtained by multiplying Hill functions to fit AND logic has practical physical significance of effective equilibrium dissociation constant. The parameter $K$ obtained by OR logical fitting by adding Hill functions up does not correspond to the actual underlying physical meaning, only with fitting significance. These results provide a theoretical reference for dealing with transcriptional regulation in the presence of multiple TFs by summating or multiplying Hill function of eq. (2).

Finally, taking LacI repressor system as an example, combining with relevant parameters measured in experiments, we deduce the *fold change* and $F_{reg}$ corresponding to this regulatory process. By fitting the two functions with Hill function of repressors in eq. (2), we not only verified the conclusions we have obtained about the cooperative regulations, but also quantitatively observed the effects of strength of promoter and strength of interactions among monomers (TFs). We found, Hill coefficient as an accurate indication only relys on strong interactions among monomers but has nothing to do with promoter condition; the fitting threshold $K$ heavily depends on promoter condition, which results in the effects from the number of RNAP and binding energy of RNAP to DNA chain on $K$ once weak promter is not met.

In conclusion, as a fitting function for transcriptional regulation, Hill function has a good fitting effect. Especially, if the underlying physical picture is not concerned, only from the perspective of phenomenological fitting function, Hill function and the "fitting logic gate function" should have broad applicable conditions. That is a reason why Hill function enjoys great popularity in gene transcriptional regulation models. However, if the reaction picture in the regulation process is expected to be read from Hill function obtained by fitting, such as obtaining the equilibrium dissociation constant and the number of TFs of related reactions, certain conditions need to be met by the system, which relates to the strength of promoters and the strength of interactions between TFs.

**Acknowledgements**

The work was supported in part by Young Talent fund of University Association for Science and Technology in Shaanxi, China (20210506), National Undergraduate Innovation and Enterpreneurship Program (202110700039) , Science and Technology Program of Gansu Province ( 21JR1RG309).

### Author Contributions

Conceived and designed the projects: W.S., Y.M. Performed analytic analysis: Y.M, W.S. N.H. Developed the computational models: Y.M., W.S., M.P. Conducted the simulation: Y.M., W.S., P. H., Y.D., Wrote the manuscript: W.S., Y.M., P. H., Y.D., D.W., Y.X.

### Conflict of interest

The authors declare that they have no conflict of interest.

## Figure legends

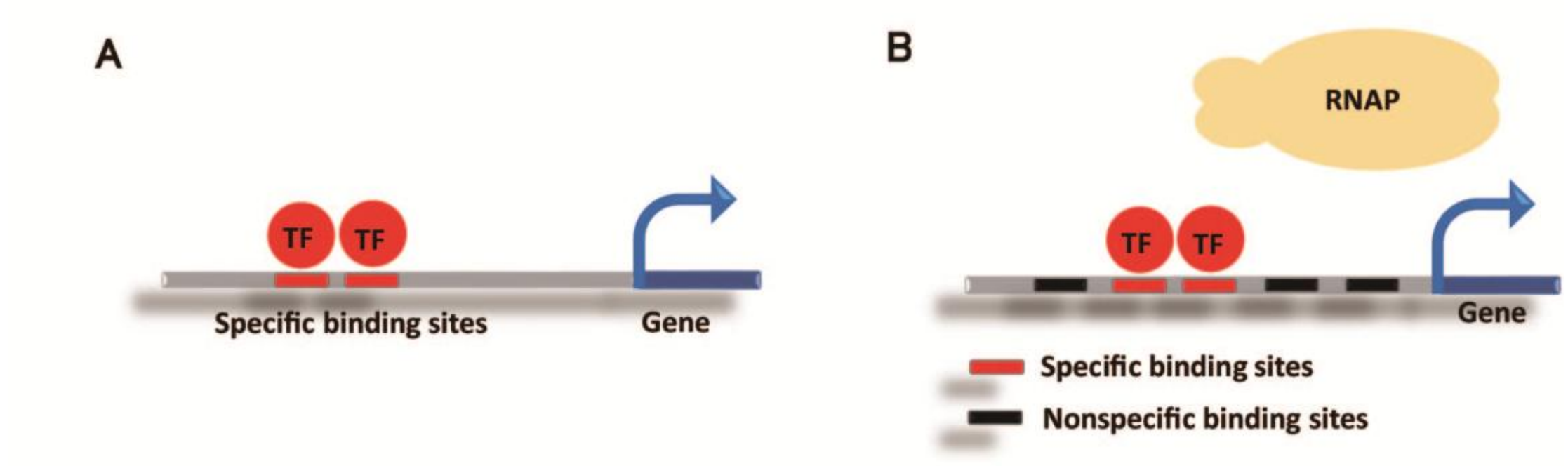

Figure 1. Schematic diagram of transcriptional regulation model. (A) Hill function model considering the binding of TFs and specific binding sites on the DNA chain. (B) Considering the interactions of TFs, RNAP and specific/nonspecific binding sites on the DNA chain.

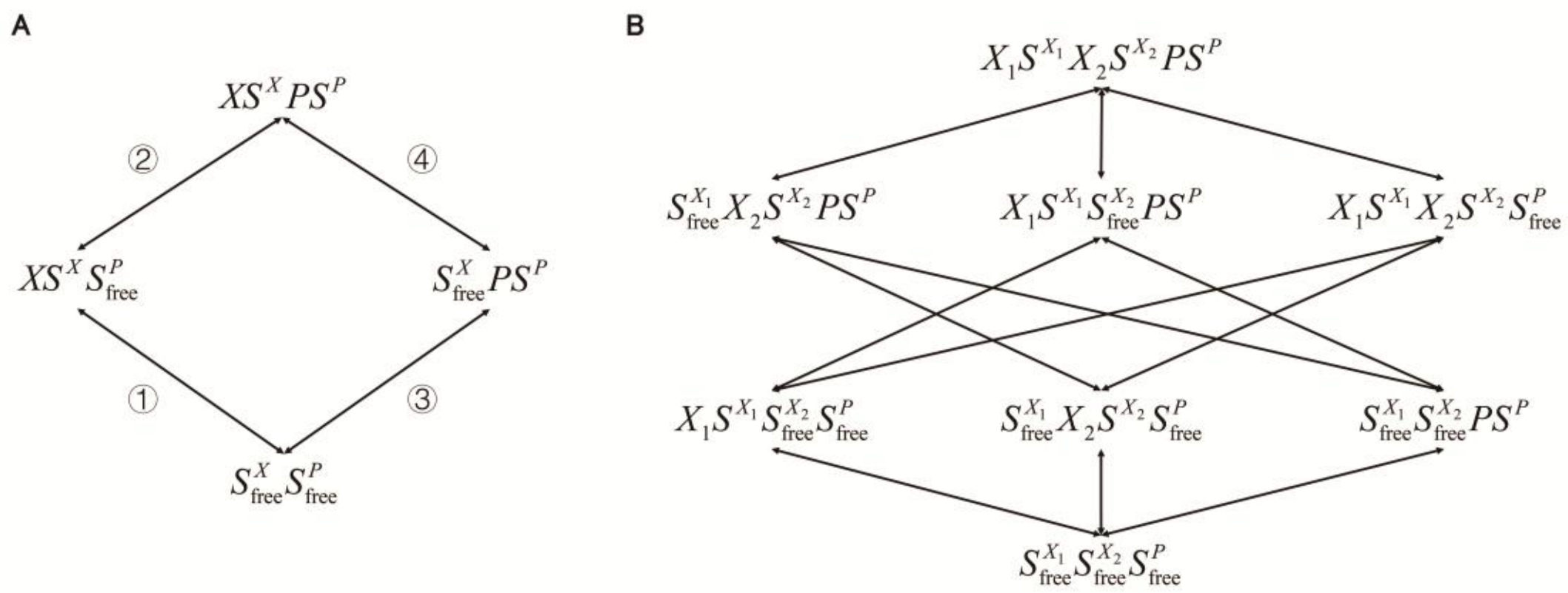

Figure 2 (A) The four occupied conformations in Table 1 and their transformations. (B) The eight occupied conformations in Table 2 and their transformations.

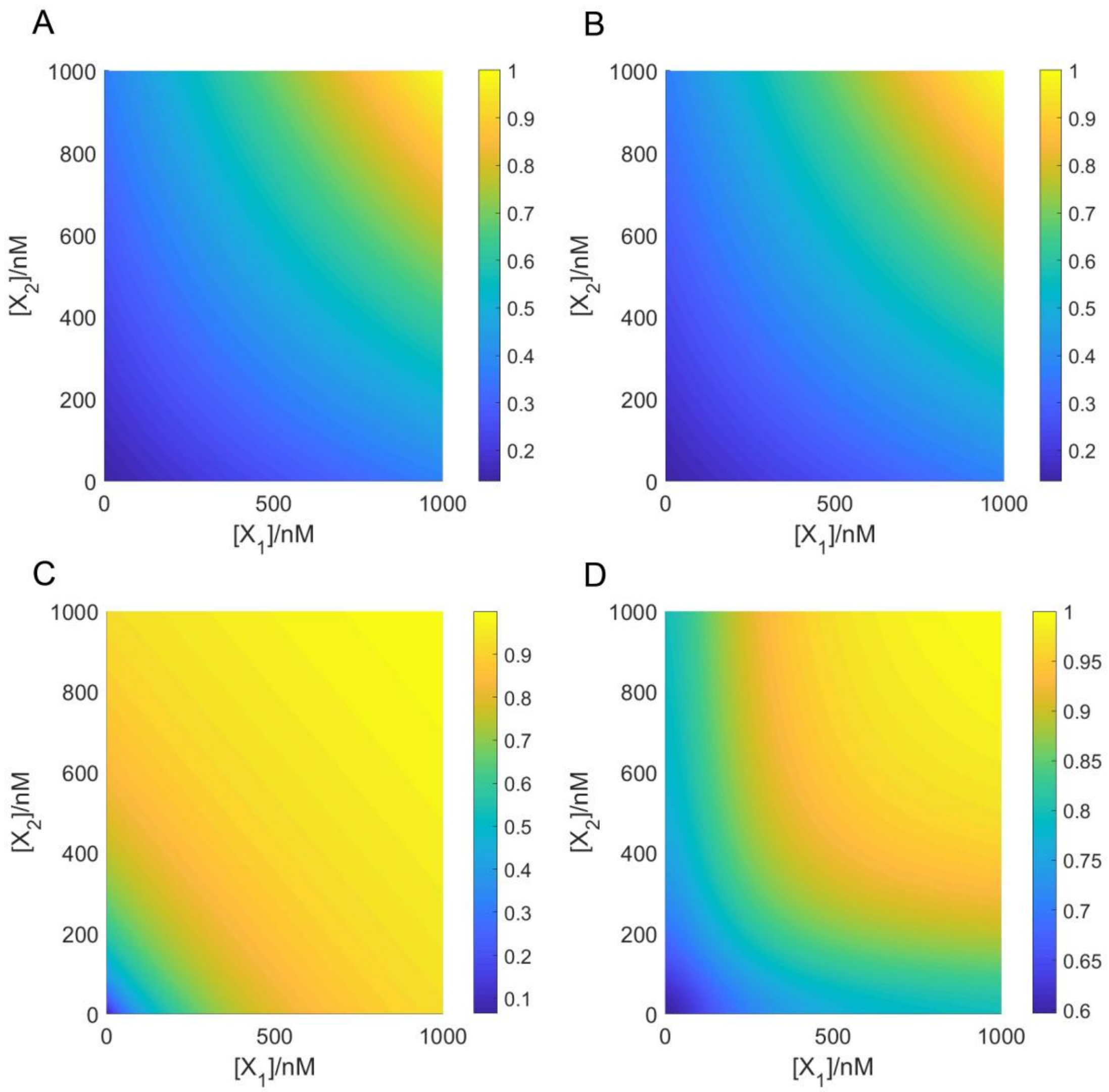

Figure 3 Regulatory factor diagram and fitting effect diagram based on "fitting logic gate function" under two kinds of logic with two co-regulated TFs. The regulatory factor diagrams are normalized, namely $\tilde{F}_{reg}([X_1],[X_2]) \equiv F_{reg}([X_1],[X_2]) / [F_{reg}([X_1],[X_2])]_{\max}$. (A) Regulatory factor diagram with two activators under AND logic in eq. (16), where $\omega_{X_1P} = \omega_{X_2P} = 20, K_{X_1} = K_{X_2} = 10000\text{nM}$ is taken. (B) The normalized fitting function $\tilde{f}_{\text{AND}}([X_1],[X_2]) \equiv f_{\text{AND}}([X_1],[X_2]) / [f_{\text{AND}}([X_1],[X_2])]_{\max}$ is used to fit $\tilde{F}_{reg}([X_1],[X_2])$ in A, and the fitting parameters are as follows: $n_i$=1, $K_i$=10000nM=$K_{Xi}$, $c_i$=0.0526 ($i$=1,2), RMES=9.2570×10$^{-12}$. (C) Regulatory factor diagram with two activators under OR logic in eq. (18), where $\omega_{X_1P} = \omega_{X_2P} = 20, K_{X_1} = K_{X_2} = 200\text{nM}$ is taken. (D) The normalized fitting function $\tilde{f}_{\text{OR}}([X_1],[X_2]) \equiv f_{\text{OR}}([X_1],[X_2]) / [f_{\text{OR}}([X_1],[X_2])]_{\max}$ is used to fit $\tilde{F}_{reg}([X_1],[X_2])$ in C, and the fitting parameters are as follows: $n_i$ =1.6608, $K_i$=241.75nM, $c_i$=1.3502 ($i$ = 1,2), RMES=0.0459.

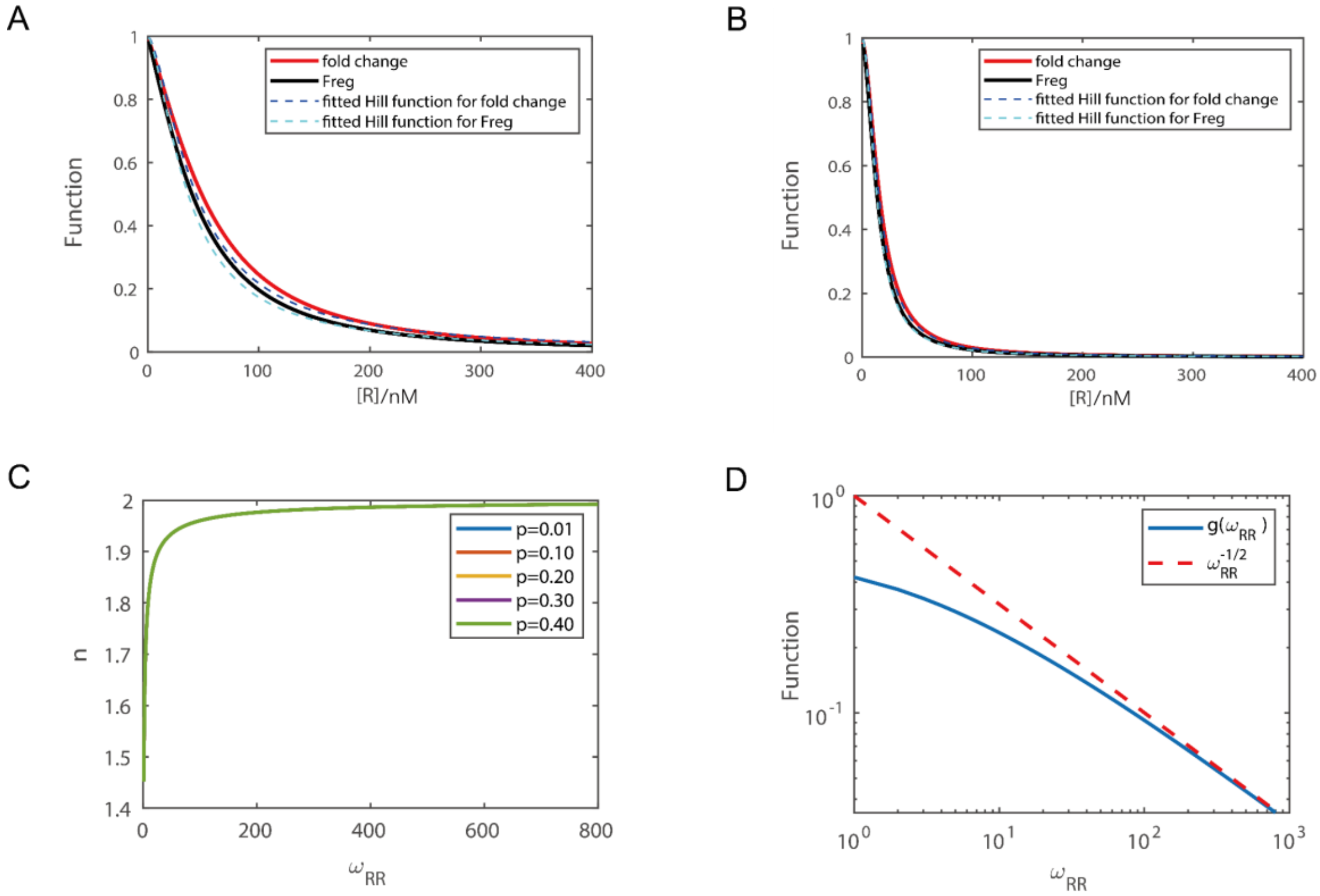

Figure 4 Regulatory response curve of LacI dimer and fitting curve of Hill function. (A-B) The horizontal axis [*R*] represents LacI dimer concentration, and the vertical axis represents *fold change* and $F_{reg}$. The solid line is the response curve based on experimental data in refs. [25] and [39] and theoretical equation. Where, $K_R = [NS]e^{\Delta\varepsilon_{rd}/K_BT}$ (see eq. (B.1) in Appendix B). $N_{NS}$ is taken 5000,000, $[N_{NS}]$=5000,000nM, and the number of RNAP *P* is taken 1000. It is approximated that $PNS \approx P$=1000, $\Delta\varepsilon_{pd}$=-7.4$k_BT$, and $\Delta\varepsilon_{rd}$=-10.5$k_BT$ in calculating *p*, resulting in *p*=0.33. The interaction factor $\omega_{RR}$ between the two LacI dimers is taken 5 (A), 800 (B). The dotted line is Hill function fitting curve. Hill coefficient *n* and regulation threshold *K* (unit: nM) are 1.54 and 43.56 (fitting parameters of *fold change* curve in fig. 4A); 1.54 and 36.25 (fitting parameters of $F_{reg}$ curve in fig. 4A); 1.99 and 5.48 (fitting parameters of *fold change* curve in fig. 4B); 1.99 and 4.75 (fitting parameters of $F_{reg}$ curve in fig. 4B). (C) The relationship between Hill coefficient *n* obtained by fitting and the cooperative effect intensity $\omega_{RR}$ of repressors at different *p* values. The curves are precisely coincident with each other when taking different *p* values. (D) The trend of functions $g(\omega_{RR})$ (solid blue line) and $\omega_{RR}^{-1/2}$ (dashed red line) as $\omega_{RR}$ changes. See the text and Appendix E for the definition of $g(\omega_{RR})$.

## Appendix A. Derivation of $F_{reg}$ in Eq. (10)

Fig. 2A includes four reversible reactions and four configurations of the DNA chains occupied by transcription factors (TFs) and RNA polymerase (RNAP). According to the quasi-equilibrium approximation, the four reversible reactions finally all reach steady states. Using the law of mass action, we can get the proportional relationship between the steady-state concentration of the corresponding four configurations, and then get the probability of RNAP occupancy on the promoter.

**Reaction 1**: $XNS + S^X_{\text{free}}S^P_{\text{free}} \rightleftharpoons XS^X S^P_{\text{free}} + NS_{\text{free}}$ (A.1)

Based on the quasi-equilibrium assumption, we can get the equilibrium dissociation constant of Reaction 1.

$$K_1 = \frac{[XNS][S^X_{\text{free}}S^P_{\text{free}}]}{[XS^X S^P_{\text{free}}][NS_{\text{free}}]} \equiv \frac{K_X}{[NS_{\text{free}}]} \approx \frac{K_X}{[NS]}$$

where the effective equilibrium dissociation constant of the TF binding to its specific binding site is defined as:

$$K_X \equiv [XNS][S^X_{\text{free}}S^P_{\text{free}}] \big/ [XS^X S^P_{\text{free}}]$$

then we get:

$$[XS^X S^P_{\text{free}}] = \frac{[XNS]}{K_X}[S^X_{\text{free}}S^P_{\text{free}}] \tag{A.2}$$

**Reaction 2:** $PNS + XS^X S^P_{\text{free}} \rightleftharpoons XS^X PS^P + NS_{\text{free}}$ (A.3)

The equilibrium dissociation constant of the reaction is:

$$K_2 = \frac{[PNS][XS^X S^P_{\text{free}}]}{[XS^X PS^P][NS_{\text{free}}]} \approx \frac{N_{PNS}}{N_{NS}} \frac{[XS^X S^P_{\text{free}}]}{[XS^X PS^P]}$$

The same approximation $N_{NS_{\text{free}}} \approx N_{NS}$ is used here and below as in the main text, then we have:

$$[XS^X PS^P] = \frac{N_{PNS}}{K_2 N_{NS}}[XS^X S^P_{\text{free}}] = \frac{[XNS]N_{PNS}}{N_{NS}K_X K_2}[S^X_{\text{free}}S^P_{\text{free}}] \tag{A.4}$$

we also use eq. (A.2) to get the second equal sign.

**Reaction 3**: $PNS + S^X_{\text{free}} S^P_{\text{free}} \rightleftharpoons S^X_{\text{free}} PS^P + NS_{\text{free}}$ (A.5)

The equilibrium dissociation constant of the reaction is:

$$K_3 = \frac{[PNS]\left[S^X_{\text{free}} S^P_{\text{free}}\right]}{\left[S^X_{\text{free}} PS^P\right][NS_{\text{free}}]} \equiv \frac{K_P}{[NS_{\text{free}}]} \approx \frac{K_P}{[NS]}$$

The effective equilibrium dissociation constant of RNAP binding to its specific binding site is defined as:

$$K_P \equiv [PNS]\left[S^X_{\text{free}} S^P_{\text{free}}\right] / \left[S^X_{\text{free}} PS^P\right].$$

As a result:

$$\left[S^X_{\text{free}} PS^P\right] = \frac{[PNS]\left[S^X_{\text{free}} S^P_{\text{free}}\right]}{[NS_{\text{free}}]K_3} \approx \frac{N_{PNS}}{N_{NS} K_3} \left[S^X_{\text{free}} S^P_{\text{free}}\right] \quad \text{(A.6)}$$

On the other hand, since TFs do not occupy their specific binding sites in this process, there is no interaction between RNAP and TFs, which reflects the intrinsic properties of association and dissociation between RNAP and promoter. Similar to eq. (5) in the main text, we have $K_3 = \exp\left(\Delta\varepsilon_{pd} / k_B T\right)$ and we can obtain the following equation by substituting it into eq. (A.6)

$$\left[S^X_{\text{free}} PS^P\right] = \frac{N_{PNS}}{N_{NS} e^{\Delta\varepsilon_{pd}/k_B T}} \left[S^X_{\text{free}} S^P_{\text{free}}\right] \quad \text{(A.7)}$$

**Reaction 4**: $XNS + S^X_{\text{free}} PS^P \rightleftharpoons XS^X PS^P + NS_{\text{free}}$ (A.8)

The equilibrium dissociation constant of the process can be obtained as:

$$K_4 = \frac{[XNS]\left[S^X_{\text{free}} PS^P\right]}{[XS^X PS^P][NS_{\text{free}}]} \equiv \frac{K'_X}{[NS_{\text{free}}]} \approx \frac{K'_X}{[NS]}$$

in which $K'_X \equiv \dfrac{[XNS]\left[S^X_{\text{free}} PS^P\right]}{[XS^X PS^P]}$.

So we can get:

$$\left[XS^X PS^P\right] = \frac{[XNS]\left[S^X_{\text{free}} PS^P\right]}{K'_X} = \frac{[XNS] N_{PNS}}{N_{NS} K'_X e^{\Delta\varepsilon_{pd}/k_B T}} \left[S^X_{\text{free}} S^P_{\text{free}}\right] \quad \text{(A.9)}$$

Here we use eq. (A.7) to get the second equal sign in eq. (A.9). It should be pointed that (A.4) is not contradictory to (A.9), and we have (see Appendix C for details) :

$$K'_X e^{\Delta\varepsilon_{pd}/k_BT} = K_2 K_X = K_X e^{\Delta\varepsilon_{pd}/k_BT} \Big/ \omega_{XP} \tag{A.10}$$

$$\left[XS^X PS^P\right] = \frac{\left[XNS\right] N_{PNS}}{N_{NS} K_X e^{\Delta\varepsilon_{pd}/k_BT}} \omega_{XP} \left[S^X_{\text{free}} S^P_{\text{free}}\right] \tag{A.11}$$

In fact, eq. (A.11) can be further transformed to a more symmetrical form:

$$\left[XS^X PS^P\right] = \frac{\left[XNS\right]\left[PNS\right]}{\left[NS\right] K_3 K_X} \omega_{XP} \left[S^X_{\text{free}} S^P_{\text{free}}\right] = \frac{\left[XNS\right]\left[PNS\right]}{K_X K_P} \omega_{XP} \left[S^X_{\text{free}} S^P_{\text{free}}\right] \tag{A.12}$$

Eq. (A.12) is helpful for us to calculate the regulatory function with two TFs in Appendix D.

By discussing the reactions above, we have got the proportional relationship between the steady-state concentration of the corresponding four configurations: (A.2), (A.7) and (A.11). We further assume that in the process of gene transcriptional regulation, the number of promoters on the DNA chain is constant, so that the sum of the concentrations of the four configurations should be a constant $\eta$.

$$\left[S^X_{\text{free}} S^P_{\text{free}}\right] + \left[XS^X S^P_{\text{free}}\right] + \left[S^X_{\text{free}} PS^P\right] + \left[XS^X PS^P\right] = \eta \tag{A.13}$$

With eqs. (A.2), (A.7), (A.11) and (A.13), we can get the concentrations of the four configurations $\left[S^X_{\text{free}} PS^P\right]$, $\left[S^X_{\text{free}} S^P_{\text{free}}\right]$, $\left[XS^X S^P_{\text{free}}\right]$, $\left[XS^X PS^P\right]$ in the quasi-equilibrium state, so that the probability of RNAP occupying the promoter can be calculated:.

$$\tilde{p}_{bound} = \left(\left[S^X_{\text{free}} PS^P\right] + \left[XS^X PS^P\right]\right) \Big/ \eta$$

From the equations above, we can obtain $\tilde{p}_{bound}$ and the corresponding regulatory factor $F_{reg}$ in the form of eq. (8) and eq. (10) respectively in the main text.

## Appendix B. Explanation about the definition of effective equilibrium dissociation constant

In the main text, we have proposed the regulatory factor $F_{reg}$. Actually this concept has been proposed by Rob Philips *et al*. in the ref. [28], where they defined it based on equilibrium statistical mechanics model of gene transcriptional regulation. In this article, we use the quasi-equilibrium assumption (this assumption is the same as the equilibrium assumption of statistical mechanics model), together with the law of mass action to obtain $F_{reg}$ in procaryotic transcriptional system. Our result should be equivalent to that of ref. [28].

Compared with the eq. (10) in main text and the third column second line of Table 1 in ref. [28], there are two differences: first, [*XNS*] in the eq. (10) represents the concentration of TF binding to nonspecific binding sites, and [*A*] in ref. [28] represents the total concentration of activators (In general, it can also be understood as the total concentration of TFs, and refers to [*X*] below). As we discussed in sect. (4) in the maintext, we concluded this slight difference can be ignored with the consideration of abundant TFs condition; Secondly, the mathematical definition of effective equilibrium dissociation constant $K_X$ is not exactly the same between this two. We can proof actually they are equivalent below.

The defination of $K_X$ in the statistical mechanics model proposed by Rob Philips *et al*. in ref. [28] is:

$$K_X = \frac{[X]}{x} = \frac{X/V}{X/N_{NS}e^{-\Delta\varepsilon_{xd}/k_BT}} = \frac{N_{NS}}{V}e^{\ln\left(K_{xd}^S/K_{xd}^{NS}\right)} = [NS]\frac{K_{xd}^S}{K_{xd}^{NS}} \tag{B.1}$$

where $x \equiv X/N_{NS}e^{-\Delta\varepsilon_{xd}/k_BT}$ and $\Delta\varepsilon_{xd} = k_BT\ln\left(K_{xd}^S/K_{xd}^{NS}\right)$ . $K_{xd}^S$ and $K_{xd}^{NS}$ are the equilibrium dissociation constants of free TFs binding to their specific and nonspecific binding sites respectively.

In eq. (B.1), $K_{xd}^S$ corresponds to the equilibrium dissociation constant of the following reaction:

$$X_{\text{free}} + S_{\text{free}}^X S_{\text{free}}^P \rightleftharpoons XS^X S_{\text{free}}^P$$

In the above reaction, the promoter is required not to be occupied by RNAP, which can eliminate the interaction factor $\varepsilon_{XP}$ and avoid changes of the equilibrium dissociation constant between TFs and the specific binding site due to the occupancy of RNAP. We believe that for TFs, RNAP is an external factor instead of intrinsic driven force of their binding and dissociation to the specific binding sites, so the influence of RNAP should be excluded when define $K_{xd}^{S}$. The equilibrium dissociation constant of the above reaction process is:

$$K_{xd}^{S} = \left[X_{\text{free}}\right]\left[S_{\text{free}}^{X} S_{\text{free}}^{P}\right] \Big/ \left[XS^{X} S_{\text{free}}^{P}\right] \tag{B.2}$$

$K_{xd}^{NS}$ in eq. (B.1) corresponds to the equilibrium dissociation constant of the following reaction:

$$X_{\text{free}} + NS_{\text{free}} \rightleftharpoons XNS$$

In the above reaction, the occupancy of the promoter by RNAP was not clearly clarified, because whether RNAP occupies the promoter or not, it doesn't have influence on the occupying of the nonspecific binding sites by TFs. Then the equilibrium dissociation constant of the corresponding process is:

$$K_{xd}^{NS} = \left[X_{\text{free}}\right]\left[NS_{\text{free}}\right] / \left[XNS\right] \tag{B.3}$$

Substituting eq. (B.2) and (B.3) into eq. (B.1), we have:

$$K_X = \left[NS\right]\frac{K_{xd}^{S}}{K_{xd}^{NS}} = \left[NS\right]\frac{\left[X_{\text{free}}\right]\left[S_{\text{free}}^{X} S_{\text{free}}^{P}\right] \Big/ \left[XS^{X} S_{\text{free}}^{P}\right]}{\left[X_{\text{free}}\right]\left[NS_{\text{free}}\right] / \left[XNS\right]} \approx \frac{\left[XNS\right]\left[S_{\text{free}}^{X} S_{\text{free}}^{P}\right]}{\left[XS^{X} S_{\text{free}}^{P}\right]}$$

In the last approximate equals sign, we applies $[NS_{\text{free}}] \approx [NS]$. So far, we have proved that the effective equilibrium dissociation constant $K_X$ defined by Rob Philips *et al* in ref. [28] is completely equivalent to that we defined in eq. (10). For other situations, like molecular $Y$, it has $K_Y$ with the similar definition mentioned above, and we won't derive later in this article.

## Appendix C. Proof of Eq. (A.10)

According to eq. (A.1) , we have:

$$\Delta\mu^0 = \mu^0_{XS^X S^P_{\text{free}}} + \mu^0_{NS_{\text{free}}} - \mu^0_{XNS} - \mu^0_{S^X_{\text{free}} S^P_{\text{free}}} = k_B T \ln K_1 \approx k_B T \ln \frac{K_X}{\left[NS\right]}$$

With eq. (A.8), we have:

$$\Delta\mu^{0\prime} = \mu^0_{XS^X PS^P} + \mu^0_{NS_{\text{free}}} - \mu^0_{XNS} - \mu^0_{S^X_{\text{free}} PS^P} = k_B T \ln K_4 \approx k_B T \ln \frac{K'_X}{\left[NS\right]}$$

Considering these two equations, we get:

$$\Delta\mu^{0\prime}-\Delta\mu^{0}=\left(\mu^{0}_{XS^{X}PS^{P}}-\mu^{0}_{XNS}-\mu^{0}_{S^{X}_{\text{free}}PS^{P}}\right)-\left(\mu^{0}_{XS^{X}S^{P}_{\text{free}}}-\mu^{0}_{XNS}-\mu^{0}_{S^{X}_{\text{free}}S^{P}_{\text{free}}}\right)=\varepsilon_{XP}=k_BT\ln\frac{K'_X}{K_X}$$

Therefore, $K_1/K_4=K_X/K'_X=\exp\left(-\varepsilon_{XP}/k_BT\right)\equiv\omega_{XP}$.

Similarly, $K_3/K_2=e^{\Delta\varepsilon_{pd}/k_BT}/K_2=\omega_{XP}$, thus we have eq. (A.10).

## Appendix D. Derivation of Eq. (12) in the maintext

Similar to (A.2), we have:

$$\left[X_1S^{X_1}S^{X_2}_{\text{free}}S^{P}_{\text{free}}\right]=\frac{\left[X_1NS\right]}{K_{X_1}}\left[S^{X_1}_{\text{free}}S^{X_2}_{\text{free}}S^{P}_{\text{free}}\right] \tag{D.1}$$

$$\left[S^{X_1}_{\text{free}}X_2S^{X_2}S^{P}_{\text{free}}\right]=\frac{\left[X_2NS\right]}{K_{X_2}}\left[S^{X_1}_{\text{free}}S^{X_2}_{\text{free}}S^{P}_{\text{free}}\right] \tag{D.2}$$

Similar to eq. (10), these two equations above can be considered as the definitions of effective equilibrium dissociation constants of TF $X_1$ and $X_2$ binding to the specific binding sites on DNA. It can be shown that this definition is equavilent to Rob Philips's definition of $K_X$ in ref. [28] (see Appendix B for details). Meanwhile, similar to eq. (A.12)[a], we can get:

$$\left[X_1S^{X_1}X_2S^{X_2}S^{P}_{\text{free}}\right]=\frac{\left[X_1NS\right]\left[X_2NS\right]\omega_{X_1X_2}}{K_{X_1}K_{X_2}}\left[S^{X_1}_{\text{free}}S^{X_2}_{\text{free}}S^{P}_{\text{free}}\right] \tag{D.3}$$

RNAP can bind to the promoters with four different configurations, and the specific reactions are:

① $PNS+S^{X_1}_{\text{free}}S^{X_2}_{\text{free}}S^{P}_{\text{free}}\rightleftharpoons S^{X_1}_{\text{free}}S^{X_2}_{\text{free}}PS^{P}+NS_{\text{free}}$

Similar to eq. (6) in the maintext, we can get $\frac{\left[S^{X_1}_{\text{free}}S^{X_2}_{\text{free}}S^{P}_{\text{free}}\right]}{\left[S^{X_1}_{\text{free}}S^{X_2}_{\text{free}}PS^{P}\right]}=\frac{N_{NS}}{N_{PNS}}e^{\Delta\varepsilon_{pd}/k_BT}$, which means:

$$\left[S^{X_1}_{\text{free}}S^{X_2}_{\text{free}}PS^{P}\right]=\frac{N_{PNS}}{N_{NS}e^{\Delta\varepsilon_{pd}/k_BT}}\left[S^{X_1}_{\text{free}}S^{X_2}_{\text{free}}S^{P}_{\text{free}}\right] \tag{D.4}$$

② $PNS+X_1S^{X_1}S^{X_2}_{\text{free}}S^{P}_{\text{free}}\rightleftharpoons X_1S^{X_1}S^{X_2}_{\text{free}}PS^{P}+NS_{\text{free}}$

---

[a] We only consider the interactions between the protein and protein/DNA , thus RNAP and TFs are identical but just with different binding energy when considering their binding with DNA, so we can regard $X$ as $X_1$, and $P$ as $X_2$ in eq. (A.12) to meet the conditions we are discussing.

Similar to eq. (A.11), we can get;

$$\left[X_1S^{X_1}S_{\text{free}}^{X_2}PS^P\right]=\frac{\left[X_1NS\right]N_{PNS}}{N_{NS}K_{X_1}e^{\Delta\varepsilon_{pd}/k_BT}}\omega_{X_1P}\left[S_{\text{free}}^{X_1}S_{\text{free}}^{X_2}S_{\text{free}}^{P}\right] \tag{D.5}$$

③ $PNS+S_{\text{free}}^{X_1}X_2S^{X_2}S_{\text{free}}^{P}\rightleftharpoons S_{\text{free}}^{X_1}X_2S^{X_2}PS^P+NS_{\text{free}}$

Similar to eq. (A.11), we can get:

$$\left[S_{\text{free}}^{X_1}X_2S_2PS^P\right]=\frac{\left[X_2NS\right]N_{PNS}}{N_{NS}K_{X_2}e^{\Delta\varepsilon_{pd}/k_BT}}\omega_{X_2P}\left[S_{\text{free}}^{X_1}S_{\text{free}}^{X_2}S_{\text{free}}^{P}\right] \tag{D.6}$$

④ $PNS+X_1S^{X_1}X_2S^{X_2}S_{\text{free}}^{P}\rightleftharpoons X_1S_1X_2S_2PS^P+NS_{\text{free}}$

Similar to eq. (D.4), and with the substitution below:

$$\left[S_{\text{free}}^{X_1}S_{\text{free}}^{X_2}S_{\text{free}}^{P}\right]\rightarrow\left[X_1S^{X_1}X_2S^{X_2}S_{\text{free}}^{P}\right]$$

$$\left[S_{\text{free}}^{X_1}S_{\text{free}}^{X_2}PS^P\right]\rightarrow\left[X_1S^{X_1}X_2S^{X_2}PS^P\right]$$

$$\Delta\varepsilon_{pd}\rightarrow\Delta\varepsilon_{pd}+\varepsilon_{X_1P}+\varepsilon_{X_2P}$$

The last substitution is due to the fact that both two TFs occupy specific binding sites, and both of them have an attractive effect on the RNAP occupied on the promoter, which is equivalent to increasing the binding energy between RNAP and the promoter by $\varepsilon_{X_1P}+\varepsilon_{X_2P}$. So we get:

$$\begin{aligned}\left[X_1S^{X_1}X_2S^{X_2}PS^P\right]&=\frac{N_{PNS}\omega_{X_1P}\omega_{X_2P}}{N_{NS}e^{\Delta\varepsilon_{pd}/k_BT}}\left[X_1S^{X_1}X_2S^{X_2}S_{\text{free}}^{P}\right]\\&=\frac{N_{PNS}}{N_{NS}e^{\Delta\varepsilon_{pd}/k_BT}}\frac{\left[X_1NS\right]\left[X_2NS\right]\omega_{X_1P}\omega_{X_2P}\omega_{X_1X_2}}{K_{X_1}K_{X_2}}\left[S_{\text{free}}^{X_1}S_{\text{free}}^{X_2}S_{\text{free}}^{P}\right]\end{aligned} \tag{D.7}$$

With eq. (D.3), we can get the second equal sign above. Similar to eq. (A.13), assuming that the number of promoters on DNA is conserved in the process of gene transcription regulation, the sum of the concentration of the eight configurations is a conserved quantity (set as $\eta$ ),

$$\begin{aligned}&\left[S_{\text{free}}^{X_1}S_{\text{free}}^{X_2}S_{\text{free}}^{P}\right]+\left[X_1S^{X_1}S_{\text{free}}^{X_2}S_{\text{free}}^{P}\right]+\left[S_{\text{free}}^{X_1}X_2S^{X_2}S_{\text{free}}^{P}\right]+\left[X_1S^{X_1}X_2S^{X_2}S_{\text{free}}^{P}\right]+\\&\left[S_{\text{free}}^{X_1}S_{\text{free}}^{X_2}PS^P\right]+\left[X_1S^{X_1}S_{\text{free}}^{X_2}PS^P\right]+\left[S_{\text{free}}^{X_1}X_2S_2PS^P\right]+\left[X_1S^{X_1}X_2S^{X_2}PS^P\right]=\eta\end{aligned} \tag{D.8}$$

Combing the eq. (D.1) to (D.8), the proportional relationship of the steady-state concentration of eight configurations mentioned above can be solved, so we can get the probability of RNAP occupying the promoter:

$$\tilde{p}_{bound} = \left( \left[ S_{\text{free}}^{X_1} S_{\text{free}}^{X_2} PS^P \right] + \left[ X_1 S^{X_1} S_{\text{free}}^{X_2} PS^P \right] + \left[ S_{\text{free}}^{X_1} X_2 S_2 PS^P \right] + \left[ X_1 S^{X_1} X_2 S^{X_2} PS^P \right] \right) \Big/ \eta$$

With simple algebraic derivation, we can get the $\tilde{p}_{bound}$ with the form of eq. (8) in the maintext, where $F_{reg}$ is given by eq. (12) in the maintext.

## Appendix E. The fitting of *fold change* with Hill function

We use Hill function of repressors as eq. (E.1) to fit the *fold change* as eq. (E.2) and $F_{reg}$ as eq. (E.3) respectively:

$$h([R]) = \frac{1}{1 + \frac{[R]^n}{K^n}} \tag{E.1}$$

$$fold\ change = \frac{1+p}{1 + p + \frac{2[R]}{K_R} + \frac{[R]^2}{K_R^2}\omega_{RR}} \tag{E.2}$$

$$F_{reg} = \frac{1}{1 + \frac{2[R]}{K_R} + \frac{[R]^2}{K_R^2}\omega_{RR}} \tag{E.3}$$

Since $F_{reg}$ can be regarded as a special case of *fold change* with $p$=0, so here we only fit *fold change* with Hill function $h([R])$ . In order to reduce independent variables, we define $r \equiv [R]/K_R$ , $k \equiv K/K_R$ to normalize $K$ and $[R]$. In this way, the *fold change* in eq. (E.2) can be transformed to:

$$fold\ change = \frac{1}{1 + \frac{2r + r^2\omega_{RR}}{1+p}} \tag{E.4}$$

and the Hill function in eq. (E.1) can be transformed to:

$$h(r) = \frac{1}{1 + \frac{r^n}{k^n}} \tag{E.5}$$

Comparing eq. (E.4) with (E.5), we need to use a power function of $(r/k)^n$ to fit a quadratic polynomial $(2r + r^2\omega_{RR})/(1+p)$ . Since $p$ only contributes to a global constant $1/(1+p)$ in the quadratic polynomial, it should only have effect on the fitted *k* rather than Hill coefficient *n* and the fitting accuracy. Then we absorb $(1+p)$ into a power function like $(r/k')^n$ to fit the quadratic polynomial $2r+r^2\omega_{RR}$, where $k' \equiv k(1+p)^{-1/n}$. We can easily found that in this condition $k'$ corresponds to the value of *k* with *p*=0, *i.e.* $k' = k(p=0)$. So there is $k(p) = k(p=0)(1+p)^{1/n}$, which means the fitting of $K/K_R(1+p)^{-1/n}$ is the function of cooperative factor $\omega_{RR}$ but has no relevance with *p* in any condition, so we set $K/K_R(1+p)^{-1/n} = g(\omega_{RR})$, *i.e* $K = K_R(1+p)^{1/n} g(\omega_{RR})$, where $g(\omega_{RR})$ is the function of $\omega_{RR}$.

$g(\omega_{RR})$ is shown with the solid blue line in fig. 4B in the maintext, which is a monotonically decreasing function of $\omega_{RR}$ in a form of $g(\omega_{RR}) \approx \omega_{RR}^{-1/2}$ (as the red dotted line in fig. 4B). We found when $\omega_{RR}$ is large, quadratic polynomials degenerate back into quadratic monomials $(2r+r^2\omega_{RR})/(1+p) \approx r^2\omega_{RR}/(1+p)$, which naturally leads *n*=2 and $k = (1+p)^{1/2}\omega_{RR}^{-1/2}$ with the fitting function $(r/k)^n$, and that's why *n* approaches to be 2 as the increase of $\omega_{RR}$ in fig. 4A in the maintext. So it's easy to have $g(\omega_{RR}) \approx \omega_{RR}^{-1/2}$ compared with the defination of *k*.